\documentclass[conference]{IEEEtran}

\newcounter{problem}
\newcounter{save@equation}
\newcounter{save@problem}
\usepackage{amsmath}

\makeatletter

\usepackage[left=1.62cm,right=1.62cm,top=1.9cm]{geometry}
\usepackage{epsfig}
\usepackage{array}
\newcolumntype{L}[1]{>{\raggedright\let\newline\\\arraybackslash\hspace{0pt}}m{#1}}
\newcolumntype{C}[1]{>{\centering\let\newline\\\arraybackslash\hspace{0pt}}m{#1}}
\newcolumntype{R}[1]{>{\raggedleft\let\newline\\\arraybackslash\hspace{0pt}}m{#1}}
\usepackage[short]{optidef}
\usepackage{amsthm} 
\usepackage{mathrsfs}
\newcommand{\bc}{\begin{center}}
\newcommand{\ec}{\end{center}}
\newcommand{\be}{\begin{equation}}
\newcommand{\ee}{\end{equation}}
\usepackage{flexisym}
\usepackage{algorithmicx}
\usepackage{algorithm}
\usepackage{siunitx}
\usepackage{comment}
\usepackage{soul}
\usepackage{algpseudocode}

\newcommand{\bnu}{\begin{enumerate}}
\newcommand{\enu}{\end{enumerate}}
\usepackage{cite}
\usepackage{url}
\usepackage[utf8]{inputenc}
\usepackage[T1]{fontenc}
\usepackage{amsmath}
\usepackage{amsmath}

\usepackage{amsfonts}
\usepackage{amssymb}
\usepackage[version=4]{mhchem}
\usepackage{stmaryrd}
\usepackage{graphicx}
\usepackage[export]{adjustbox}
\graphicspath{ {./images/} }
\usepackage{breqn}
\usepackage{lipsum}
\usepackage{mathtools}
\usepackage{caption}
\usepackage{float}
\usepackage{subcaption}
\usepackage{soul}
\usepackage{blindtext, graphicx}
\usepackage{cuted}
\usepackage{color}

\usepackage{amsfonts}
\usepackage{multicol}
\usepackage{lipsum}
\usepackage{cuted}

\usepackage{graphicx}
\graphicspath{ {./images/} }

\newtheoremstyle{case}{}{}{}{}{}{:}{ }{}

\ifCLASSINFOpdf
\else
\fi

\begin{document}
\title{EcoEdgeTwin: Enhanced 6G Network via Mobile Edge Computing and Digital Twin Integration}
\author{Synthia Hossain Karobi$^\dagger$, Shakil Ahmed$^\ddagger$, Saifur Rahman Sabuj$^*$, and Ashfaq Khokhar$^\ddagger$\\
$^\dagger$Department of Electrical and Computer Engineering, University of California Riverside, USA.\\
$^\ddagger$Department of Electrical and Computer Engineering, Iowa State University, USA.\\
$^*$Department of Electrical and Electronic Engineering, BRAC University, Bangladesh.
}

\maketitle
\vspace{-0.5 cm}
\begin{abstract}
In the 6G era, integrating Mobile Edge Computing (MEC) and Digital Twin (DT) technologies presents a transformative approach to enhance network performance through predictive, adaptive control for energy-efficient, low-latency communication. This paper presents the EcoEdgeTwin model, an innovative framework that harnesses the synergy between MEC and DT technologies to ensure efficient network operation. 
We optimize the utility function within the EcoEdgeTwin model to balance enhancing users' Quality of Experience (QoE) and minimizing latency and energy consumption at edge servers. This approach ensures efficient and adaptable network operations, utilizing DT to synchronize and integrate real-time data seamlessly.
 Our framework achieves this by implementing robust mechanisms for task offloading, service caching, and cost-effective service migration. Additionally, it manages energy consumption related to task processing, communication, and the influence of DT predictions, all essential for optimizing latency and minimizing energy usage. 
 Through the utility model, we also prioritize QoE, fostering a user-centric approach to network management that balances network efficiency with user satisfaction. A cornerstone of our approach is integrating the advantage actor-critic algorithm, marking a pioneering use of deep reinforcement learning for dynamic network management. This strategy addresses challenges in service mobility and network variability, ensuring optimal network performance matrices. Our extensive simulations demonstrate that compared to benchmark models lacking DT integration,  EcoEdgeTwin framework significantly reduces energy usage and latency while enhancing QoE.
\end{abstract}
\vspace{-0.5 cm}
\section{Introduction}
As we edge closer to the dawn of the 6G  networks, we witness a paradigm shift in the swiftly evolving telecommunications landscape. This imminent phase in network evolution transcends the capabilities of 5G, catering to the escalating demand for higher data throughput, reduced latency, and more sophisticated connectivity solutions. In the crucible of this transformation, Mobile Edge Computing (MEC) stands as a pivotal element, poised to fulfill the rigorous performance requisites and host intelligent services for an extensively connected world, all underpinned by Artificial Intelligence (AI) techniques.
Amidst the progress, the deployment of these networks encounters notable hurdles, such as dynamic network conditions, energy efficiency, latency sensitivity, user mobility, and service migration cost, etc., which pose a formidable challenge in forecasting network dynamics and crafting effective offloading strategies \cite{allioui2023exploring}. 

In response to these complexities, Digital Twin (DT) technology emerges as a beacon of innovation, enabling a virtual mirroring of physical assets, i.e., MEC, within a digital realm \cite{do2022digital}. 
The synergy between MEC and DT paves the way for a transformative era in the 6G digital infrastructure, optimizing decision-making processes through AI and enhancing the symbiosis between physical and virtual realms.
Current research in MEC predominantly focuses on balancing computation latency and energy consumption without adequately addressing the complexities introduced by user mobility and evolving network dynamics expected in 6G environments \cite{zhang2022drl, ndikumana2018joint}. 
This gap in MEC demands DT technology integration for real-time and predictive network management. Despite DT's potential to mirror and digitally predict the state of physical assets, its application in MEC for enhancing operational efficiency still needs to be explored. Adopting Deep Reinforcement Learning (DRL) may provide a pivotal opportunity to address this shortfall \cite{zhang2022drl}. DRL stands out for its ability to learn optimal policies through trial and error, making it ideal for complex, variable environments like MEC. Incorporating DRL, specifically the Advantage Actor-Critic (A2C) algorithm,  can significantly improve decision-making processes, leveraging DT insights for a proactive and adaptive network response \cite{zhu2020parallel}. A2C optimizes network operations by framing decisions as actions in a reinforcement learning setup, improving performance and user satisfaction. This approach promises to bridge the current research gap by offering a robust solution for dynamic task offloading and resource management.

\vspace{-0.1 cm}
\textit{Related Work:} Recent advancements in MEC have led to significant research on optimizing network resources and energy consumption. A pivotal area of focus has been the integration of DT technology with MEC to enhance operational efficiency and Quality of Experience (QoE). 
For instance, the authors in \cite{do2022digital} explored the potential of DTs in MEC environments, emphasizing their role in achieving real-time network synchronization and predictive analytics for better resource allocation.
The heuristic algorithm also underscored the interplay between communication, computation, and service migration to maximize cost-effectiveness in\cite{dinh2016offloading}.
Similarly, the authors in \cite{wu2021digital} discussed how DT could mirror physical network components, offering insights into optimizing task offloading decisions based on user behavior.
The authors in \cite{zhang2016energy} proposed an efficient online dynamic mobile offloading scheme rooted in DRL and introduced the DT-enabled edge network architecture to predict the future state of the MEC environment.
DTs were heralded as a critical enabler in the forthcoming 6G era, promising to support through predictive analytics and intelligent operations in \cite{ndikumana2018joint}.
While these contributions underscore the evolving landscape of MEC and DT research, realizing these technologies' full potential within a cohesive framework remains a burgeoning field of study, necessitating further exploration and innovation.
\begin{figure}[ht]
\centering
\includegraphics[width=3.2in]{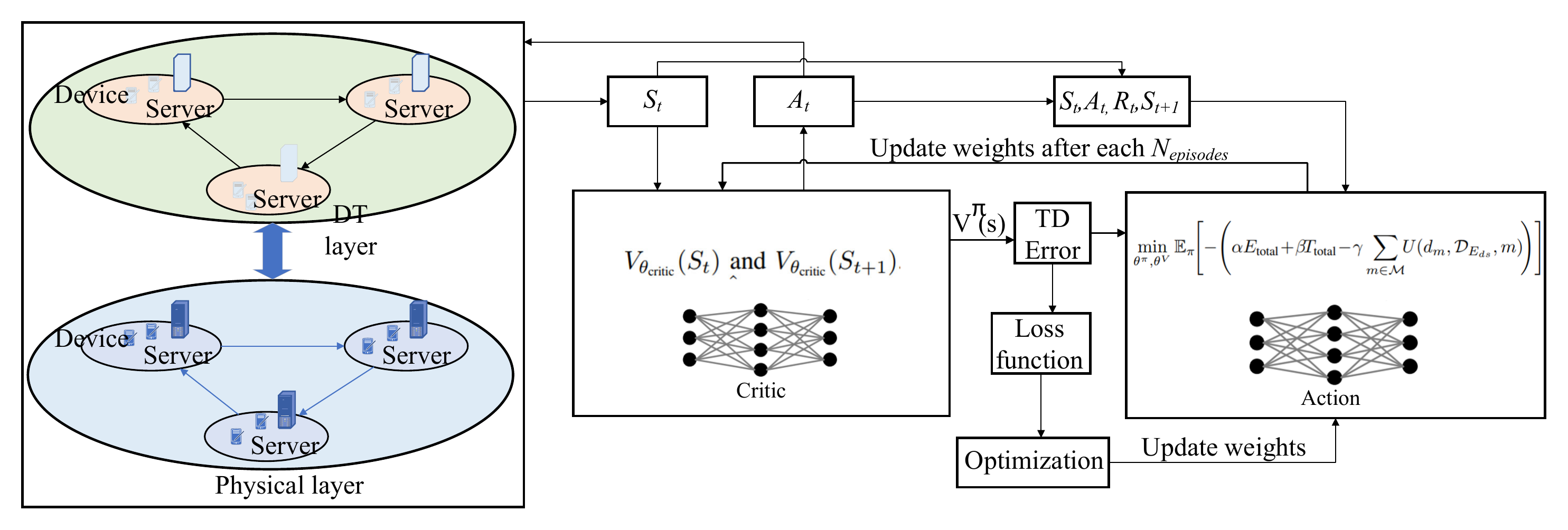}
\caption{EcoEdgeTwin model}
\label{Fig_SM}
\end{figure}

\textit{Contribution:} The contributions of this paper are:
\textit{1)} We develop the EcoEdgeTwin model that synergizes MEC with DT for real-time synchronization of network components to significantly improve energy efficiency and reduce latency by ensuring QoE.
\textit{2)} We implement robust mechanisms such as task offloading, service caching, and cost-effective service migration. These strategies are crucial for minimizing latency and energy consumption.
\textit{3)} We elaborate on an energy model by capturing task processing, communications, and the influence of DT predictions within MEC environments to support sustainable network functionality.
\textit{4)} Our utility model aligns network operations with user satisfaction metrics, focusing on a user-centric network management approach that balances QoE with latency and energy usage.
\textit{5)} By integrating the A2C algorithm, we introduce DRL for dynamic management of task offloading and energy conservation, adeptly addressing challenges posed by user mobility and network condition variability.
\textit{6)} Extensive simulations validate our model's effectiveness, showing it outperforms traditional benchmark frameworks by reducing energy consumption and latency while enhancing QoE.

\textit{Notation:} Rest of the paper, subscripts $j$, $i$,  and  $\{ij\}$ denote entities related to user $M_j$, edge server $E_{ds_i}$, and  $M_j$-$E_{ds_i}$ association, respectively, and symbol $\mathcal{D}_{[.]}$ is  DT-related entities.

\section{Porposed EcoEdgeTwin Model}
This section discusses the EcoEdgeTwin model, as follows:

\textit{Physical Infrastructure Layer:}
The physical layer of the EcoEdgeTwin network comprises edge servers, denoted by $\mathcal{E}$,  designed to manage offloaded computational tasks. 
We denote the set of edge servers as $\mathcal{E} = \{E_{ds_1}, E_{ds_2}, \ldots, E_{ds_n}\}$, where $E_{ds_i}$ represents the $i$-th edge server in the system for $i \in {1, 2, \ldots, n}$, and $n$ is the total count of edge servers deployed across the network. We assume that edge servers are ideally placed across the network to serve various users.  Let be the set of all mobile users as $\mathcal{M} = \{M_1, M_2, \ldots, M_u\}$,
where $M_{j}$, for $j \in {1, 2, \ldots, u}$,   is a user within the network.
$M_{j}$ has computation capabilities to execute tasks locally. However, due to the limited computation capabilities to execute computation-intense applications, $M_{j}$ can wirelessly offload its tasks to $E_{ds_i}$ with an additional cost of latency and energy consumption.

\textit{DT Integration Layer:}
DT integration layer, $\mathcal{D}$, complements the physical infrastructure by adding a digital dimension to the edge servers and the users, thus enhancing decision-making and operational efficiency.
The digital layer of the EcoEdgeTwin network comprises edge servers, denoted by $\mathcal{D}_{\mathcal{E}}$,  designed to manage offloaded computational tasks with low latency. 
We define $\mathcal{D}_{\mathcal{E}}= \{\mathcal{D}_{E_{ds_1}}, \mathcal{D}_{E_{ds_2}}, \ldots, \mathcal{D}_{E_{ds_n}}\}$,
where $\mathcal{D}_{E_{ds_i}}$, for $i \in {1, 2, \ldots, n}$. Let us denote the set of all mobile users as $\mathcal{D}_{\mathcal{M}} = \{\mathcal{D}_{M_1}, \mathcal{D}_{M_2}, \ldots, \mathcal{D}_{M_u}\}$,
where $\mathcal{D}_{M_{j}}$, for $j \in {1, 2, \ldots, u}$,   is a single user within DT network.

\subsubsection{Task Offloading}
The task offloading mechanism is crucial in determining the optimal execution server for computational tasks generated by users \cite{9691475}. This decision-making process evaluates each task's computational demand (CPU cycles and data size) against the processing capabilities of the available edge server, $E_{ds_i}$. 
Within the network, each task $\mathcal{O}_{ij}$ from user ${M_{j}}$ is characterized by a tuple $\{D_{j}, C_{et_{j}},L_j,M_{\text{mig}_j}\}$, where  $D_j$ is the data size of the task from $M_j$, $C_{et_{j}}$ is the required computational resource i.e., CPU cycles, $L_j$ is  the latency, and $M_{\text{mig}_j}$ is the migration cost of user $M_j$'s offloaded task $T_{ij}$ moving from one server to another.
We define data size, $D_j$, as follows: 
\begin{equation}
\label{Eq_Dj}
D_j=\underbrace {\alpha_j D_j}_{\text{Local}} + \underbrace {\bigg [ T_{ij}=\sum_{E_{ds_i} \in \mathcal{E}} \pi_{ij} \beta_{ij} D_j \bigg ]}_{\text{Offloaded}} 
\end{equation} 
where  $\boldsymbol{\alpha}
 = \{\alpha_j\}_{\forall j}$ is the amount of the task processed locally.
 We define the offloaded task as $T_{ij}$ in (\ref{Eq_Dj}), where $\boldsymbol{\beta}
 = \{\beta_{ij}\}_{\forall i,j}$ is the offloading factor of the $M_j$ to $E_{ds_i}$, satisfying $0 \leq \alpha_j \leq 1$, $0 \leq \beta_{ij} \leq 1$.
and 
$\alpha_j + \sum_{E_{ds_i} \in \mathcal{E}} \pi_{ij} \beta_{ij} = 1$. $\pi_{ij}$ is the binary offloading decision variable indicating whether user $M_j$ is associated with edge server $E_{ds_i}$. 
 The required computational resource, $C_{et_{j}}$ is expressed a follows: 
\begin{equation}
C_{et_{j}}=\alpha_j C_{et_{j}} + \sum_{E_{ds_i} \in \mathcal{E}} \pi_{ij} \beta_{ij} C_{et_{j}}
\end{equation}

When the offloaded task $T_{ij}$ moves, service migration is subsequently required, and the user transfers from one edge server to another while maintaining a steady QoE. Let $C_f$ be the fixed cost required for service migration and $h_{j}$ binary indicator. Migration cost on whether $M_j$ migrates is defined as follows: 
\begin{equation}
M_{\text{mig}_j}=\sum_{M_j \in \mathcal{M}} h_{j} T_{ij} C_f
\end{equation}
where 
\begin{equation}
h_{j} = 
\begin{cases}
1, & \!\!\!\!\!\text{if } M_j \text{ migrates to a new edge server }, \\
0, & \!\!\!\!\! \text{otherwise}.
\end{cases}
\end{equation}

The DT services replicate the physical infrastructure, we, therefore, define that task in DT, $\mathcal{O}_{ij}(\mathcal{D}_{D_{j}}, \mathcal{D}_{C_{et_{j}}},\mathcal{D}_{L_j},\mathcal{D}_{M_{\text{mig}_j})}$, originated from $\mathcal{D}_{M_j}$  using DT layer as follows:
data size, $\mathcal{D}_{D_j}=\alpha_j D_j + \sum_{\mathcal{D}_{E_{ds_i}} \in \mathcal{D}_{\mathcal{E}}} \pi_{ij} \beta_{ij} \mathcal{D}_{D_j}$, the computational resource, $\mathcal{D}_{C_{et_{j}}}=\alpha_j\mathcal{D}_{C_{et_{j}}} + \sum_{\mathcal{D}_{E_{ds_i}} \in \mathcal{E}} \pi_{ij} \beta_{ij} \mathcal{D}_{C_{et_{j}}}$, and the migration cost, $\mathcal{D}_{M_{\text{mig}_j}}=\sum_{\mathcal{D}_{M_j} \in \mathcal{D}_{\mathcal{M}}} h_{j}  \mathcal{D}_{C_f}T_{ij}$, where $\mathcal{D}_{C_f}=C_f+\tilde{C}_f$. $\tilde{C}_f$ is the discrepancy from physical to DT infrastructure.
\subsubsection{Service Caching Strategy}
Optimal service caching reduces latency and ensures minimal computational overload by dynamically and periodically caching data and computational outcomes on the edge servers. We define a caching scheme to combine user response behavior   for optimal caching design for physical and DT as follows:
\begin{equation}
\label{Eq_SCS}
 D_{j},\mathcal{D}_{D_{j}}  \leq M_{\text{mig}_j}, \forall E_{ds_i}, \forall M_j,\forall \mathcal{D}_{E_{ds_i}}, \forall \mathcal{D}_{M_j} 
\end{equation}
\subsubsection{Communication Model}
We denote the channel coefficients between mobile user \( M_j \) and edge server \( E_{ds_i} \) by \( \rho_{ij} \), which encapsulates both large-scale and small-scale fading effects, represented as \( \rho_{ij} = \sqrt{\gamma_{ij} s_{ij}} \), where \( \gamma_{ij} \) is the path-loss coefficient and \( s_{ij} \) is the small-scale fading coefficient for the channel between \( M_j \) and \( E_{ds_i} \).
The achievable transmission rate for the task offloaded from \( M_j \) to \( E_{ds_i} \) is modeled as:
\begin{equation}
R_{ij} = B \log_2 \left( 1 + \frac{\pi_{ij} p_j |\rho_{{ij}}|^2}{ d_{ij} \sigma^2_i} \right),
\end{equation}
where \( p_j \) is the transmit power of \( M_j \), \( \sigma^2_i \) is the noise variance at $E_{ds_i}$ and $d_{ij}$ is the distance between $M_j$ and $E_{ds_i}$. $B$ is channel bandwidth.
\subsubsection{Latency}  
The task \( \mathcal{O}_{ij} \) associated with user \( M_j \) incurs a latency \( L_j \), which is composed of local computing latency \( L_j^{\text{l}} \) and edge computing latency from \( M_j \) to \( E_{ds_i} \), denoted as \( L_{ij} \).
The local processing latency is given by: $L_j^{\text{l}} = \frac{\alpha_j C_{et_j}}{f_j}$,
where \( \alpha_j \) is the portion of the task processed locally with the processing rate, \( f_j \). \( C_{et_j} \) denotes the required computation resource.
The edge processing latency is estimated as:
 $L_{ij} = \frac{\pi_{ij} \beta_{ij} C_{et_j}}{R_{ij} f_i} $,
where \( \pi_{ij} \) is the user association indicator, \( \beta_{ij} \) is the offloading factor of the \( M_j \)  to \( E_{ds_i} \).
Assuming that the deviation of the CPU processing frequency between the physical and their DT counterparts can be acquired in advance, the computing latency gap between the actual value and the DT estimation for local processing is  $\Delta L_j^{\text{l}} = \frac{\alpha_j C_{et_j} \tilde{f}_j}{f_j (f_j - \tilde{f}_j)} $, where \( \tilde{f}_j \) represents the estimated processing capability of \( M_j \) in DT.
Similarly, the latency gap for edge processing is:
  $\Delta L_{ij} = \frac{\pi_{ij} \beta_{ij} C_{et_j} \tilde{f}_i}{R_{ij} f_i (f_i - \tilde{f}_i)} $, where \( \tilde{f}_i \) represents the estimated processing capability of \( E_{ds_i} \) in DT.
Thus, the total latency due to the physical layer and DT for task \( \mathcal{O}_{ij} \) by \( M_j \) to \( E_{ds_i} \) can be expressed as: 
\begin{equation}
    L_j = L_j^{\text{l}} + \Delta L_j^{\text{l}} + L_{ij} + \Delta L_{ij}
\end{equation}

\subsubsection{Energy Model}
We define the energy consumed by edge servers due to processing tasks, communication between nodes, and the indirect influence of  DT predictions on these energy expenditures as follows:
\begin{equation}
\begin{aligned}
&E_{ij} =  \!\!\!\sum_{E_{ds_i} \in \mathcal{E}} \!\! \left( \sum \!\! \frac{\lambda_{T_{ij}} \cdot e_{\text{unit}}}{f_{i} \!+ \!\tilde{f}_i} + E_{\text{comm}}(E_{ds_i}, M_j) \!\! \right)  \\
& \frac{\alpha_jC_{et_j}}{f_{j} \!+ \!\tilde{f}_j}+ \sum_{M_j \in \mathcal{M}} \left( E^{\text{up}}_{j} + E^{\text{down}}_{j} \right) + \sum_{\mathcal{D}_{M_j} \in \mathcal{D}_{\mathcal{M}}} \mathcal{D}_{E_{\text{DT}}},
\end{aligned}
\end{equation}
where $\lambda_{T_{ij}} $ is  the computational workload associated with task ${T_{ij}}$. $f_{i}$ is the computational capacity (frequency) of the edge server at $E_{ds_i}$, $e_{\text{unit}}$ is the energy consumption per unit of computation and $\tilde{f}_j$ is the discrepancy in computational capability predicted by the DT for edge server $E_{ds_i}$. $\mathcal{D}_{E_{\text{DT}}}$ represents the energy consumed due to operations and predictions made by the DT. $E^{\text{up}}_{j}$ and $ E^{\text{down}}_{j}$ represent the energy consumed for uploading and downloading data for user $M_j$, adjusted based on insights from the user-specific DT. $E_{\text{comm}}(E_{ds_i}, M_j)$ is the the energy consumed for communication by edge server 
$E_{ds_i}$ with all users in the set $\mathcal{M}$.

\subsubsection{ DT Discrepancies}
To effectively manage the variances between the computational capabilities of physical edge servers and their  DT representations, we introduce a discrepancy factor, $G(E_{ds_i}, M_j,\mathcal{D}_{E_{ds_i}}, \mathcal{D}_{M_j})$, for each task $T_{ij}$ offloaded by user $M_j$ to edge server $E_{ds_i}$. This factor is defined as:
\begin{equation}
    G_{ij}  = -\frac{\lambda_{T_{ij}}  (\hat{f}_{i} - f_{i})}{f_{i}(f_{i} + \hat{f}_{i})},
\end{equation}
where $\lambda_{T_{ij}}$ denotes the computational requirements of offloaded task $T_{ij}$, $f_{i}$ represents the actual CPU frequency of edge server $E_{ds_i}$, and $\hat{f}_{i}$ indicates the estimated CPU frequency as perceived by the  DT for user $M_j$. This discrepancy factor $G$ quantifies the impact of the DT's estimation error on the computation latency for task $T_{ij}$.

\subsubsection{QoE as Utility Function}
QoE within the EcoEdgeTwin model  captures the satisfaction concerning latency, computation efficiency, and cost savings of $M_j$ as follows:
\begin{equation}
QoE_j = \lambda^w_j \cdot W_j(L_j) + \lambda^s_j \bigg[ S_j(B_j, C_{st_j})= \frac{B_j - C_{st_j}}{B_j} \bigg ],
\end{equation}
where  
\( \lambda^w_j \) and \( \lambda^s_j \) are the weighting parameters for user satisfaction and cost savings, respectively, with \( \lambda^w_j + \lambda^s_j = 1 \).
\( S_j(B_j, C_{et_j}) \) is the cost savings score based on the user's budget, where \( C_{st_j}\) and \( B_j \) are the cost incurred due to consumption and budget of $M_j$'s computing resources. This score reflects resource utilization efficiency relative to the user's budget.
 \( W_j(L_j) \) is the user satisfaction score based on latency \( L_j \) in both physical and DT, mapped onto a mean opinion score scale with thresholds \( L_{\min} \) and \( L_{\max} \) and expressed as:
\begin{equation}
W_j(L_j) = \begin{cases} 
1, & \text{if } L_j \leq L_{\min}, \\
\frac{L_{\max} - L_j}{L_{\max} - L_{\min}}, & \text{if } L_{\min} < L_j \leq L_{\max}, \\
W_{b_j}, & \text{otherwise},
\end{cases}
\end{equation}
where  \( W_{b_j} \) denotes the baseline satisfaction score for unsatisfactory service latency.

\subsubsection{Optimization Problem}
The optimization problem stated as follows balances latency, energy consumption, and utility function:
\begin{subequations}
\begin{align}
    \min \quad & \!\!\!\!(w_{1} \cdot L_{j} + w_{2} \cdot E_{ij}) - w_{3} \!\!\! \sum_{M_j \in \mathcal{M}} U(QoE_j) \label{opt:objective_simplified} \\
    \!\!\! \text{s.t.} \quad & D_{j},\mathcal{D}_{D_{j}}  \leq M_{\text{mig}_j}, \forall E_{ds_i}, \forall M_j,\forall \mathcal{D}_{E_{ds_i}}, \forall \mathcal{D}_{M_j} , \label{opt:caching_constraint_simplified} \\
        & \!\!\!\!\!\!\!\!\!\!\!\!\!\! \mathcal{O}_{m} \!\! \leq  \! \mathcal{O}_{ij}(D_{j},\!\mathcal{D}_{D_{j}},\! C_{et_{j}}, \!\mathcal{D}_{C_{et_{j}}},\! L_j,\mathcal{D}_{L_j},\!M_{\text{mig}_j},\! \mathcal{D}_{M_{\text{mig}_j}}\!),\!\! \label{opt:offloading_strategy} \\
    & \!\!\!\!\!\!\!\!\!\!\!\!\!\! G_{ij}  = -\frac{\lambda_{T_{ij}}  (\hat{f}_{i} - f_{i})}{f_{i}(f_{i} + \hat{f}_{i})} \label{opt:DT} 
\end{align}
\end{subequations}
where objective function in (\ref{opt:objective_simplified}) minimizes total energy consumption, including computational and communicative aspects, and total latency while maximizing utility function. 
$w_{1-3}$ is the weighting factor for energy and latency minimization and utility function maximization, respectively.
$L_{j}$ is the latency, $E_{ij}$ is aggregate energy consumption.
The caching strategies within the physical infrastructure and the influence of DT are encapsulated in (\ref{opt:caching_constraint_simplified}).
The constraint of the task is captured in (\ref{opt:offloading_strategy}), where $\mathcal{O}_{m}$ is the minimum task for $M_j$.
The discrepancy between physical and DT is captured in (\ref{opt:DT}).

\section{DRL-Solution Approach}
\textit{1) System State Representation and Decision-making:}
MEC environments' dynamic and unpredictable nature, particularly regarding user mobility and variable network conditions, necessitates a sophisticated model for optimizing task offloading decisions. Our approach incorporates an A2C framework enriched by insights from DT to address the issue. 

\textbf{State Space, $S_t$:} The state encapsulates the network's physical attributes and the digital twins' predictive insights, accurately reflecting the operational context in real-time. It optimizes task offloading and service caching strategies to minimize latency and energy consumption while maximizing QoE for users. We define this state as: 
$  S_t =  \{D_{j},\!\mathcal{D}_{D_{j}},\! C_{et_{j}}, \!\mathcal{D}_{C_{et_{j}}},\! L_j,\mathcal{D}_{L_j},\!M_{\text{mig}_j},\! \mathcal{D}_{M_{\text{mig}_j}}, {f}_i, d_{ij}, \tilde{f}_i, {f}_j, \tilde{f}_j,$ $ T_{ij},\mid E_{ds_i} \in \mathcal{E}, M_j \in \mathcal{D}_{\mathcal{M}}, \mathcal{D}_{\mathcal{E}}, \mathcal{D}_{M_j} \in \mathcal{D}_{\mathcal{M}}\}$,
where $d_{ij}$ denotes the distance between user $M_j$ and edge server $E_{ds_i}$, $f_{i}$ signifies the computational capability of $E_{ds_i}$, $\tilde{f}_i$ indicates discrepancies in DT's performance estimation for $E_{ds_i}$, $M_{\text{mig}_j}$ represents the service migration cost for user $M_j$ to $E_{ds_i}$, and $\mathcal{D}_{E_{ds_i}}, \mathcal{D}_{M_j}$ embody DT insights for edge servers and users, respectively. 

\textbf{Action Space, $A_t$:} In the EcoEdgeTwin model, actions encompass decisions on task offloading and service caching that are pivotal for enhancing the network's operational efficiency. These decisions are intricately linked to optimizing physical infrastructure and digital insights provided by Digital Twins. The action set is broadened to include the dynamic interaction between physical edge servers and their digital counterparts, defined as:
$      A_t = \{\text{offload to } E_{ds_i}, \text{ cache at } E_{ds_i}, 
 \text{ DT adjustments } | E_{ds_i} \in \mathcal{E}, \mathcal{D}_{E_{ds_i}}\}$,  
where actions involve not only selecting an edge server $E_{ds_i}$ for task offloading and determining the optimal caching strategy within the set of physical edge servers $\mathcal{E}$ but also making adjustments based on DT, $\mathcal{D}_{E_{ds_i}}$ insights. 

\textbf{Reward Function, $R_t$:} The reward function quantitatively evaluates the impact of decisions on network operations. It balances minimizing latency and energy usage against maximizing QoE. The reward function is formalized as:
\begin{equation}
R_t = -(w_{1}\cdot L_{j} +w_{2} \cdot E_{ij} ) - w_{3} \sum_{M_j \in \mathcal{M}} U(QoE_j),
\end{equation}
where  $w_{1}$, $w_{2}$, and $w_{3}$ prioritize energy efficiency, latency reduction, and utility function, i.e., QoE.

\begin{algorithm}
\caption{Actor-Critic Training for EcoEdgeTwin}
\label{alg:ecoedgetwin_a2c}
\textbf{Input:}  Learning rates: $\alpha_{\text{actor}}$ for the actor, $\alpha_{\text{critic}}$ for the critic, training episodes: $N_{\text{episodes}}$, maximum steps per episode: $T_{\text{max}}$.
\textbf{Initialization:} $\theta_{\text{actor}}$, $\theta_{\text{critic}}$ for actor and critic networks.
\begin{algorithmic}[1]
\For{$\text{episodes} = 1, \ldots, N_{\text{episodes}}$}
    \State Initialize environment to obtain initial state $S_1$.
    \For{$t = 1, \ldots, T_{\text{max}}$}
        \State Selects $A_t$ based on policy $\pi_{\theta_{\text{actor}}}(A_t | S_t)$.
        \State Execute $A_t$, observe reward $R_{t+1}$, new state $S_{t+1}$.
        \State Critic calculates value $V_{\theta_{\text{critic}}}(S_t)$ and $V_{\theta_{\text{critic}}}(S_{t+1})$.
        \State Compute advantage estimate $\hat{A}_t = R_{t+1} + w_{3} V_{\theta_{\text{critic}}}(S_{t+1}) - V_{\theta_{\text{critic}}}(S_t)$.
        \State Update $\theta_{\text{actor}}$ using gradient ascent: $\theta_{\text{actor}} \leftarrow \theta_{\text{actor}} + \alpha_{\text{actor}} \nabla_{\theta_{\text{actor}}} \log \pi_{\theta_{\text{actor}}}(A_t | S_t) \hat{A}_t$.
        \State Update $\theta_{\text{critic}}$ using gradient descent on squared loss: $\theta_{\text{critic}} \leftarrow \theta_{\text{critic}} - \alpha_{\text{critic}} \nabla_{\theta_{\text{critic}}} (\hat{A}_t)^2$.
        \State Update the system state $S_t \leftarrow S_{t+1}$.
    \EndFor
\EndFor
\end{algorithmic}
\textbf{Return} Optimized parameters $\theta_{\text{actor}}$, $\theta_{\text{critic}}$.
\end{algorithm}

\textit{2) A2C Algorithm in  EcoEdgeTwin:} This framework integrates DTs for edge servers and the system with the MEC environment to create a real-time digital representation of the operational context. The EcoEdgeTwin model utilizes the A2C algorithm to optimize task offloading and energy management, responding adaptively to user mobility and changing network conditions. The A2C framework, with its dual components of actor and critic networks, facilitates informed decision-making by leveraging state representations that combine physical network attributes with digital twin insights.
$S_t$ amalgamates real-time data from the physical layer and predictive insights from digital twins as:
$  S_t =  \{D_{j},\!\mathcal{D}_{D_{j}},\! C_{et_{j}}, \!\mathcal{D}_{C_{et_{j}}},\! L_j,\mathcal{D}_{L_j},\!M_{\text{mig}_j},\! \mathcal{D}_{M_{\text{mig}_j}}, {f}_i, d_{ij}, \tilde{f}_i, {f}_j, \tilde{f}_j,$ $ T_{ij},\mid E_{ds_i} \in \mathcal{E}, M_j \in \mathcal{D}_{\mathcal{M}}, \mathcal{D}_{\mathcal{E}}, \mathcal{D}_{M_j} \in \mathcal{D}_{\mathcal{M}}\}$,
where $d_{ij}$ signifies the proximity between user $M_j$ and edge server $E_{ds_i}$, affecting decision-making related to data transmission and task offloading. $f_{i}$ and $\hat{f}_{i}$ represent the actual and estimated computational capabilities of $E_{ds_i}$, respectively, influencing computations related to task processing capabilities and offloading strategies. $M_{\text{mig}_j}$ denotes the cost of dynamically migrating tasks across servers to optimize network responsiveness and resource utilization.
The Actor-network, $\pi(A_t | S_t; \theta^\pi)$, parameterized by $\theta^\pi$, suggests actionable strategies based on $S_t$, minimizing latency and energy usage while maximizing user satisfaction. Actions $A_t$ are chosen to optimize $\mathcal{O}_{ij}$ and service caching  decisions:
$A_t = \{\text{offload } T_{ij} \text{ to } E_{ds_i}, \!\! \text{ cache } \!\! D_{j} \!\! \text{ at } E_{ds_i} \! \mid E_{ds_i} \in \mathcal{E}\}.$
Critic network, with parameters $\theta^V$, assesses the potential value of state $S_t$ towards achieving the model's objectives, aiding in refining the Actor's policy through feedback on the expected rewards as follows:
\begin{equation}
    \!\! \min_{\theta^\pi, \theta^V} \mathbb{E}_{\pi}\!\! \left[ -\!\left( \! w_{1} \cdot  L_{j}+ w_{2} \cdot E_{ij}  \!-\! w_{3} \sum_{M_j \in \mathcal{M}}\!\! U(QoE_j) \!\right) \right]\!\!, \!\!
\end{equation}
where $E_{ij}$ and $L_{j}$ denote the system-wide energy consumption and latency, respectively. $U(QoE_j)$ quantifies user satisfaction based on service delay and digital twin insights. $w_{1-3}$ serve as balancing coefficients.
The trained model parameters are used as input, and then decisions regarding offloading are made in iterations over a given number of episodes and time steps. For these concluding decisions, the policy of the actor-network guides proactive actions that consider the system's current state based on physical and digital twin status knowledge. Such actions are designed for a decision space that optimizes a given objective function, considering latency, energy consumption, and service migration cost. Algorithm~\ref{alg:a2c_task_offloading} uses A2C to optimize task offloading proactively. It traverses through episodes, further segmented into steps over which decisions for task offloading or service caching are defined. The actor network plays designed actions concerning the current observation in the environment, whereas the critic network assesses each action's value by approximating the temporal difference (TD) error. 
\begin{algorithm}
\caption{A2C Algorithm for Task Offloading}
\label{alg:a2c_task_offloading}
\textbf{Input:}  $\alpha_{\text{actor}}$, $\alpha_{\text{critic}}$, $\theta_{\text{actor}}$, $\theta_{\text{critic}}$

\textbf{Initialization:} Initialize parameters $\theta_{\text{actor}}$, $\theta_{\text{critic}}$ randomly.
\begin{algorithmic}[1]
\For{each episode}
    \State Initialize the digital twin state representation $S_0$.
    \For{each step $t$ within the episode}
        \State Compute action probabilities $\pi(A_t | S_t, \theta_{\text{actor}})$.
        \State Sample action $A_t$ based on $\pi(A_t | S_t)$.
        \State Execute $A_t$, transitioning to $S_{t+1}$ and $R_{t+1}$.
        \State Compute TD error $\delta_t = R_{t+1} + w_{3} V(S_{t+1}, \theta_{\text{critic}}) - V(S_t, \theta_{\text{critic}})$.
        \State Update critic by minimizing loss: $\mathcal{L}(\theta_{\text{critic}}) = \delta_t^2$.
        \State Update actor: $\nabla_{\theta_{\text{actor}}} \log \pi(A_t | S_t, \theta_{\text{actor}}) \cdot \delta_t$.
    \EndFor
    \If{convergence criteria met}
        \State Break
    \EndIf
\EndFor
\end{algorithmic}
\textbf{return} Optimized parameters $\theta_{\text{actor}}^*$, $\theta_{\text{critic}}^*$.
\end{algorithm}

\section{Simulation Results}
We configured parameters to mirror the operational conditions of MEC and DT environments. The simulated network operates on a transmission channel bandwidth of 20 MHz with edge servers. User devices have a transmit power that varies from 0.2 to 0.6 watts, functioning within a noise power landscape of $2 \times 10^{-12}$ watts. The tasks assigned to users require a data volume between 600 to 800 kilobytes, with a processing demand of 200 to 400 CPU cycles per kilobyte of data. The environment also incorporates a router queuing latency of 2 milliseconds and a task request latency ranging from 150 to 250 milliseconds. An integral part of our simulation is the mean estimation error of the DT's performance, set at 0.5.
The configuration for creating edge servers is defined by an area side length of 3 kilometers, encapsulating a total area of 9 square kilometers. Each edge server covers a radius of 0.15 kilometers. Within this area, the edge server density is established at 5 per square kilometer, resulting in 45 edge servers spread evenly across the simulation landscape. This setup encapsulates a dense urban environment where effective edge computing is paramount, facilitating the investigation of offloading strategies within the 6G network framework.
As the user trajectory Microsoft T drive dataset was used \cite{10.1145/1869790.1869807}, and for the A2C, two neural networks represent the Actor and Critic, each with three hidden layers using the ReLU activation function. 
T-Drive dataset encapsulates GPS trajectories from 10,357 taxis. i.e., mobile users over a week in Beijing, China, mobile users amassing about 15 million data points covering 9 million kilometers. Each taxi's trajectory data is recorded every 177 seconds, translating to an average distance interval of 623 meters. This dataset can significantly enhance mobile edge computing simulations, especially for testing task offloading strategies. The approach aids in balancing network loads, reducing latency, and improving service delivery across urban environments, leveraging the trajectories to model the movement of mobile devices and their interaction with distributed network resources. The actor loss function focused on maximizing the expected reward, encouraging actions that lead to more favorable states and outcomes. The reward function is central to the learning process, which quantifies the desirability of the agent's actions.

\begin{figure}
    \centering
    \begin{subfigure}[H]{0.20\textwidth}
        \centering
        \includegraphics[width=\textwidth]{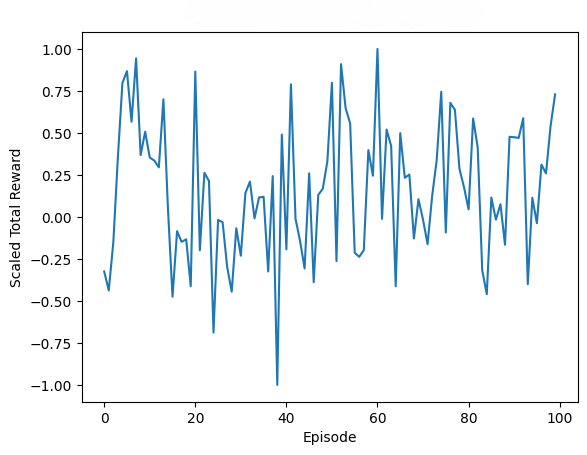}
        \caption{Scaled reward/episode}
        \label{Fig_Cov}
    \end{subfigure}
    \hfill
    \begin{subfigure}[H]{0.24\textwidth}
        \centering
        \includegraphics[width=\textwidth]{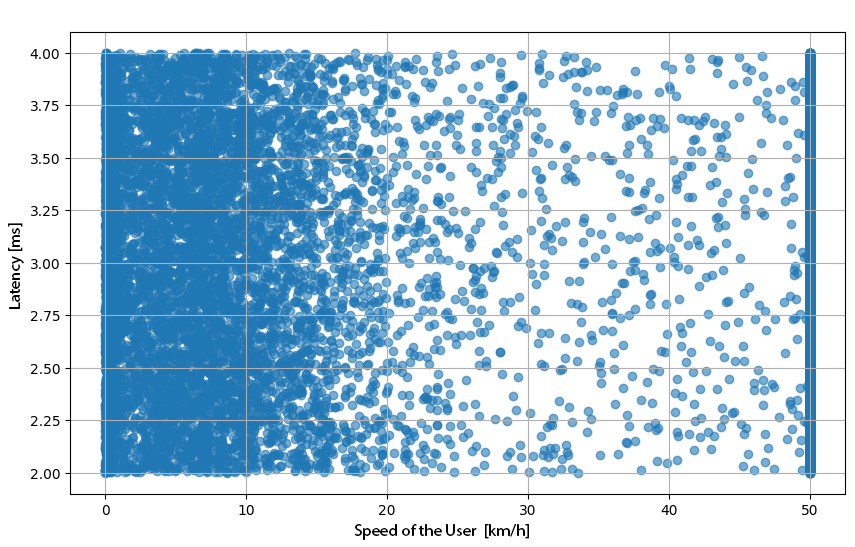}
        \caption{latency vs. user speed}
        \label{fig:throughput_extreme_conditions}
    \end{subfigure}
\caption{Performance metrics of EcoEdgeTwin model}
\label{fig:network_dynamics_examination}
\end{figure}

Fig.~\ref{Fig_Cov} illustrates the scaled total reward per episode obtained during the training of the A2C algorithm within the EcoEdgeTwin framework. Each episode is associated with one agent-environment interaction, and the scaled total reward reveals how well the Actor-Critic agent made a trade-off between energy consumption, latency, and user velocity in the offloading process. The reward is scaled to ease interpretation for visualization and explanation and support a stable training progression by maintaining the same scale of the reward axis across every episode. Fig.~\ref{Fig_Cov} also depicts a large variability in the total rewards during the initial stages of the training, highlighting high levels of exploration of the action space and a subsequent learning process. The variability in total rewards from one episode to another indicates a complex problem and active learning, with the actor-critic yet to find a stable and optimal policy. This pattern represents the adaptive nature of the EcoEdgeTwin model as it seeks to learn the best strategies for dealing with the MEC environment in real time.
The illustration in Fig.~\ref{fig:throughput_extreme_conditions} depicts the correlation between global offloading latency and the user speed in kilometers per hour. Each data point on the scatter plot represents a pair of measurements, providing insights into how the user's speed impacts the global latency when offloading tasks to edge servers. It is intuitive to assume that higher speed results in increased latency due to  more frequent server changes. 

\begin{figure}
    \centering
    \begin{subfigure}[H]{0.24\textwidth}
        \centering
        \includegraphics[width=\textwidth]{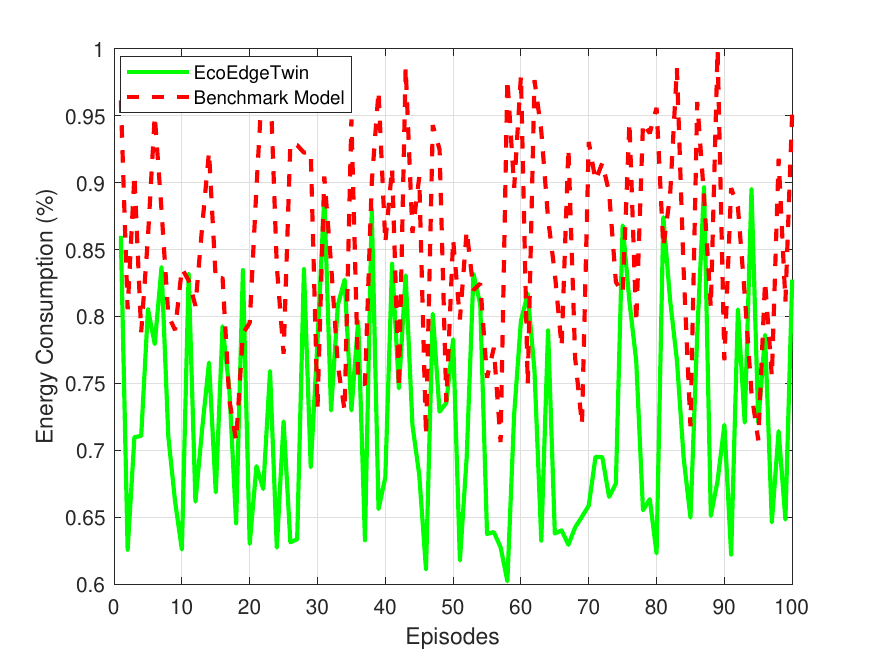}
        \caption{Energy consumption}
        \label{Fig_Energy}
    \end{subfigure}
    \hfill
    \begin{subfigure}[H]{0.24\textwidth}
        \centering
        \includegraphics[width=\textwidth]{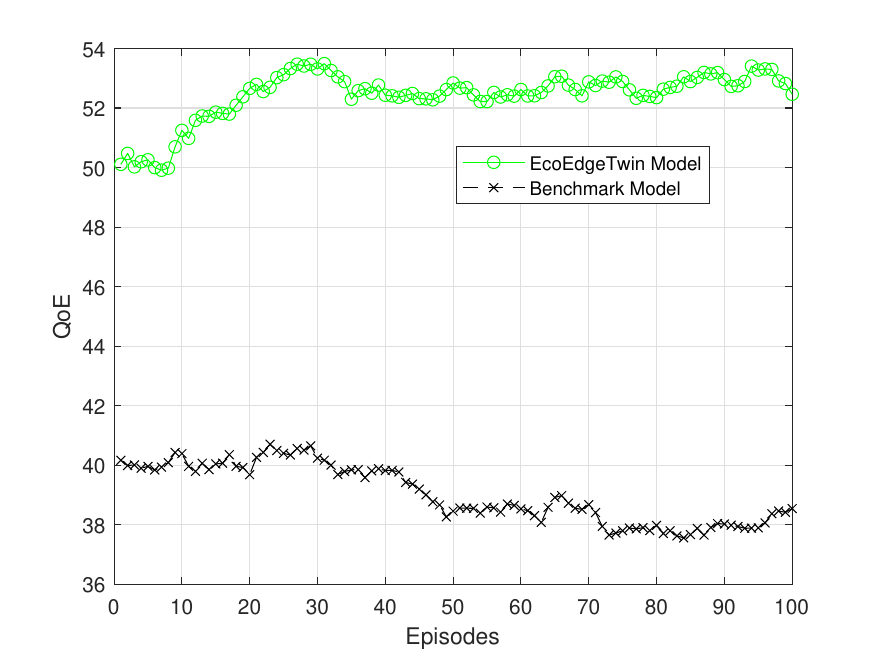}
        \caption{QoE}
        \label{fig:QoE}
    \end{subfigure}
\caption{Comparison: EcoEdgeTwin vs. benchmark}
\label{fig:EchoEdgeTwin}
\end{figure}

Fig.~\ref{Fig_Energy} shows a breakdown of energy consumption over several episodes for the EcoEdgeTwin and Benchmark models, both spanning 100 episodes. The benchmark model is a conventional network optimization strategy that lacks the integrated DT framework, relying on standard offloading and resource allocation methods.
EcoEdgeTwin model boasts lower and more consistent energy consumption percentages, indicating its impressive energy conservation capabilities compared to the Benchmark model compared to the benchmark model.
Fig.~\ref{fig:QoE} showcases the QoE metric for both models throughout 100 episodes. EcoEdgeTwin model displays consistently higher QoE values that improve over time, indicating exceptional user satisfaction. 
Fig.~\ref{fig:EchoEdgeTwin} suggests that the EcoEdgeTwin model surpasses the Benchmark model in both energy efficiency and QoE, highlighting the model's effectiveness in optimizing network performance while ensuring user satisfaction.

\section{Conclusion}
This paper presented the EcoEdgeTwin model, an innovative framework that harnessed the synergy between MEC and DT technologies to ensure efficient network operation. We optimized the utility function within the EcoEdgeTwin model to balance enhancing users' QoE and minimizing latency and energy consumption at edge servers to ensure efficient and adaptable network operations, utilizing DT to synchronize and integrate real-time data seamlessly. Our framework achieved this by implementing robust mechanisms for task offloading, service caching, and cost-effective service migration. Additionally, it managed energy consumption related to task processing, communication, and the influence of DT predictions, all essential for optimizing latency and minimizing energy usage. Through the utility model, we prioritized QoE, fostering a user-centric approach to network management that balanced network efficiency with user satisfaction. We integrated the advantage actor-critic algorithm, marking a pioneering use of DRL for dynamic network management. This strategy addressed challenges in service mobility and network variability, ensuring optimal network performance matrices. Our simulations showed that, compared to benchmark models, EcoEdgeTwin framework significantly reduced energy usage and latency while enhancing QoE compared to the benchmark model.



\begin{thebibliography}{10}
\providecommand{\url}[1]{#1}
\csname url@samestyle\endcsname
\providecommand{\newblock}{\relax}
\providecommand{\bibinfo}[2]{#2}
\providecommand{\BIBentrySTDinterwordspacing}{\spaceskip=0pt\relax}
\providecommand{\BIBentryALTinterwordstretchfactor}{4}
\providecommand{\BIBentryALTinterwordspacing}{\spaceskip=\fontdimen2\font plus
\BIBentryALTinterwordstretchfactor\fontdimen3\font minus \fontdimen4\font\relax}
\providecommand{\BIBforeignlanguage}[2]{{%
\expandafter\ifx\csname l@#1\endcsname\relax
\typeout{** WARNING: IEEEtran.bst: No hyphenation pattern has been}%
\typeout{** loaded for the language `#1'. Using the pattern for}%
\typeout{** the default language instead.}%
\else
\language=\csname l@#1\endcsname
\fi
#2}}
\providecommand{\BIBdecl}{\relax}
\BIBdecl

\bibitem{allioui2023exploring}
H.~Allioui and Y.~Mourdi, ``Exploring the full potentials of {IoT} for better financial growth and stability: {A} comprehensive survey,'' \emph{Sensors}, vol.~23, no.~19, p. 8015, 2023.

\bibitem{do2022digital}
T.~Do-Duy, D.~Van~Huynh, O.~A. Dobre, B.~Canberk, and T.~Q. Duong, ``Digital twin-aided intelligent offloading with edge selection in mobile edge computing,'' \emph{IEEE Wireless Communications Letters}, vol.~11, no.~4, pp. 806--810, 2022.

\bibitem{zhang2022drl}
S.~Zhang, H.~Gu, K.~Chi, L.~Huang, K.~Yu, and S.~Mumtaz, ``{DRL}-based partial offloading for maximizing sum computation rate of wireless powered mobile edge computing network,'' \emph{IEEE Transactions on Wireless Communications}, vol.~21, no.~12, pp. 10\,934--10\,948, 2022.

\bibitem{ndikumana2018joint}
A.~Ndikumana, N.~H. Tran, T.~M. Ho, Z.~Han, W.~Saad, D.~Niyato, and C.~S. Hong, ``Joint cache assignment and task offloading for mobile edge computing in dense networks,'' in \emph{Proc. IEEE INFOCOM}, Honolulu, HI, USA, 2018, pp. 1--9.

\bibitem{zhu2020parallel}
X.~Zhu and Y.~Du, ``A parallel approach to advantage actor critic in deep reinforcement learning.''\hskip 1em plus 0.5em minus 0.4em\relax Springer, 2020, pp. 320--327.

\bibitem{dinh2016offloading}
T.~Q. Dinh \emph{et~al.}, ``Offloading in mobile edge computing: {T}ask allocation and computational frequency scaling,'' \emph{IEEE Trans. Comput.}, vol.~65, no.~8, pp. 2419--2431, Aug. 2016.

\bibitem{wu2021digital}
Y.~Wu, K.~Zhang, and Y.~Zhang, ``Digital twin networks: {A} survey,'' \emph{IEEE Internet of Things Journal}, vol.~8, no.~18, pp. 13\,789--13\,804, 2021.

\bibitem{zhang2016energy}
Z.~Zhang \emph{et~al.}, ``Energy-efficient offloading for mobile edge computing in {5G} heterogeneous networks,'' \emph{IEEE Access}, pp. 5896--5907, 2016.

\bibitem{9691475}
T.~Do-Duy, D.~Van~Huynh, O.~A. Dobre, B.~Canberk, and T.~Q. Duong, ``Digital twin-aided intelligent offloading with edge selection in mobile edge computing,'' \emph{IEEE Wireless Communications Letters}, vol.~11, no.~4, pp. 806--810, 2022.

\bibitem{10.1145/1869790.1869807}
J.~Yuan \emph{et~al.}, ``T-drive:{ D}riving directions based on taxi trajectories,'' in \emph{Proceedings of the 18th International Conference on Advances in Geographic Information Systems}, New York, NY, USA, 2010, p. 99–108.

\end{thebibliography}
\end{document}